\documentclass{amsart}
\usepackage{amssymb}
\begin{document}
\newtheorem{theorem}{Theorem}[section]
\newtheorem{lemma}[theorem]{Lemma}
\newtheorem{remark}[theorem]{Remark}
\newtheorem{definition}[theorem]{Definition}
\newtheorem{corollary}[theorem]{Corollary}
\font\pbglie=eufm10
\def\BB{{\mathcal{B}}}
\def\cc{{\bar c}}
\def\ffrac#1#2{{\textstyle\frac{#1}{#2}}}
\def\Id{{\operatorname{Id}}}
\def\KK{\mathfrak{K}}
\def\PSpan{{\operatorname{Span}}}
\def\PP{\tilde{\mathcal{P}}} \def\qedbox{\hbox{$\rlap{$\sqcap$}\sqcup$}}
\def\QQ{\tilde{\mathcal{Q}}}
\def\SS{S}
\def\Tr{{\operatorname{Tr}}}
\def\vol{{\operatorname{vol}}} \def\weight{\operatorname{weight}}
\def\xxe{{\text{\pbglie e}}}
\def\xxi{{\text{\pbglie i}}}
\def\xxI{{\text{\pbglie I}}}
\makeatletter
 \renewcommand{\theequation}{%
 \thesection.\alph{equation}}
 \@addtoreset{equation}{section}
 \makeatother
\title[Invariance theory]{Divergence terms in the supertrace
heat asymptotics for the de Rham complex on a manifold with boundary}
\author[Gilkey et. al.]{ P. Gilkey${}^{1,2,\alpha,\beta}$, K. Kirsten${}^{2,\beta,\gamma}$,
and D. Vassilevich${}^{2,\beta}$} 
\thanks{${}^1$Research partially supported by the Mittag-Leffler Institute (Stockholm, Sweden)}
\thanks{${}^2$Research partially supported by the MPI (Leipzig, Germany)}
\thanks{${}^\alpha$Mathematics Department, University of Oregon, Eugene Or 97403 USA}
\thanks{${}^\beta$Max-Planck-Institute for Mathematics in the Sciences, Inselstrasse 22, 04103 Leipzig Germany}
\thanks{${}^\gamma$ Department of Mathematics, Baylor University, Waco, TX 76798 USA}
\begin{email}{gilkey@darkwing.uoregon.edu, klaus.kirsten@mis.mpg.de, \newline \phantom{.}
\qquad Klaus\_Kirsten@baylor.edu, vassil@itp.uni-leipzig.de} \end{email} \begin{abstract}
We use invariance theory to determine the coefficient $a_{m+1,m}^{d+\delta}$ in the super trace for the twisted
de Rham complex with absolute boundary conditions.\end{abstract}  
\maketitle
\section{Introduction}
Let $(M,g)$ be a compact Riemannian manifold of dimension $m$ with
smooth, non-empty boundary $\partial M$. Let $\phi\in C^\infty(M)$
be an auxiliary smooth function called the dilaton. Let
$d_\phi:=e^{-\phi}de^\phi$ and let $\delta_{\phi,g}:=e^\phi
\delta_g e^{-\phi}$ be the twisted exterior derivative and the
co-derivative, respectively, on the space of smooth differential
forms. The {\it twisted} or {\it Witten} Laplacian is given by:
$$\Delta_{\phi,g}^p:=d_\phi\delta_{\phi,g}+\delta_{\phi,g}d_\phi
\quad\text{on}\quad C^\infty(\Lambda^p(M))\,.$$ This operator
appears in the study of quantum $p$ form fields interacting with a
background dilaton \cite{GKVZ02,VZ00}. It has also been used in
supersymmetric quantum mechanics \cite{ABI} and in Morse theory
\cite{Wit82}.

We impose {\it absolute boundary conditions} $\BB_a$, see
\cite{G94} for details. Let $\Delta^p_{\phi,g,\BB_a}$ be the
associated realization. We need not consider relative boundary
conditions $\BB_r$ as the Hodge $\star$ operator intertwines
$\Delta_{\phi,g,\BB_a}^p$ and
$\Delta_{-\phi,g,\BB_r}^{\phantom{..}m-p}$ if $M$ is orientable
\cite{GKVZ02}. These boundary conditions are motivated by the
Hodge-de Rham theorem which shows
$$\ker(\Delta^p_{\phi,g,\BB_a})=H^p(M)\,.$$

The fundamental solution $e^{-t\Delta_{\phi,g,\BB_a}^p}$ of the
heat equation is an infinitely smoothing operator which is of
trace class. Let $f\in C^\infty(M)$ be a smooth smearing function.
Work of Greiner and Seeley \cite{Gri71,Se69b} shows there is a
complete asymptotic expansion
$$\Tr_{L^2}(fe^{-t\Delta_{\phi,g,\BB_a}^p})
 \sim\textstyle\sum_{n\ge0}a_{n,m}(f,\Delta_{\phi,g}^p,\BB_a)t^{(n-m)/2}\quad\text{as}\quad
t\downarrow0\,.
$$
The {\it heat trace invariants} $a_{n,m}(\cdot)$ are locally computable. Let $\nabla_{e_m}^kf$ be the $k^{th}$ covariant derivative of $f$ with respect to the inward unit normal $e_m$ on $\partial M$. Let $dx$ and $dy$ be the Riemannian measures on $M$ and on $\partial M$, respectively. There exist local invariants $a_{n,m}(x,\Delta_{\phi ,g}^p)$ and $a_{n,m,k}(y,\Delta_{\phi ,g }^p,\BB_a)$ so that \begin{eqnarray*} a_{n,m}(f, \Delta_{\phi, g}^p,\BB_a) &=&\textstyle\int_Mf(x)a_{n,m}(x,\Delta_{\phi,g}^p)dx\\
&+&\textstyle\sum_k\int_{\partial M}\nabla_{e_m}^kf(y)\cdot
a_{n,m,k}(y,\Delta_{\phi,g}^p,\BB_a)dy\,.
\end{eqnarray*}

The interior invariants vanish if $n$ is odd; the boundary
invariants are generically non-zero for all $n\ge1$. The presence
of the smearing function $f$ localizes the problem and permits the
recovery of divergence terms which would otherwise be lost. The
presence of terms involving $\nabla_{e_m}^kf$ shows the kernel
function for the fundamental solution of the heat equation behaves
asymptotically like a distribution near the boundary as
$t\downarrow0$ . Define the local {\it supertrace heat
asymptotics} by setting: \begin{eqnarray*}
&&a_{n,m}^{d+\delta}(\phi,g)(x):=\textstyle\textstyle\sum_p(-1)^p
 a_{n,m}(x,\Delta_{\phi,g}^p),\\ &&a_{n,m,k}^{d+\delta}(\phi,g)(y):=\textstyle\textstyle\sum_p(-1)^p
 a_{n,m,k}(y,\Delta_{\phi,g}^p,\BB_a).
\end{eqnarray*}

Let $\chi(M)$ be the Euler-Poincar\'e characteristic of $M$. If
$f=1$ and if $\phi$ satisfies Neumann boundary conditions, then
\cite{GKVZ02}
$$\textstyle\textstyle\sum_p(-1)^p\Tr_{L^2}(e^{-t\Delta_{\phi,g,\BB_a}^p})=\chi(M)\,.$$
Equating terms in the asymptotic series yields:
\begin{equation}\label{eqn1.a}
\textstyle\int_Ma_{n,m}^{d+\delta}(\phi,g)(x)dx+
 \textstyle\int_{\partial M} a_{n,m,0}^{d+\delta}(\phi,g)(y)dy=\left\{\begin{array}{ll}\chi(M)&\text{if }n=m,\\0&\text{if }n\ne m\,. \end{array}\right. \end{equation}

The local index density has been computed in this setting
\cite{GKVZ02}. Let indices $i,j,...$ range from $1$ to $m$ and
index a local orthonormal frame for the tangent bundle of $M$. Let
$R_{ijkl}$ be the associated components of the Riemann curvature
tensor with the sign convention that $R_{1221}=+1$ on the unit
sphere $S^2\subset\mathbb{R}^3$. Near the boundary, normalize the
choice of the orthonormal frame so $e_m$ is the inward unit
geodesic normal. Let indices $a,b,...$ range from $1$ to $m-1$ and
index the induced orthonormal frame for the tangent bundle of the
boundary. Let $L_{ab}$ be the components of the second fundamental
form.

We adopt the Einstein convention and sum over repeated indices. Let $$\varepsilon_U^V:= g(e_{u_1}\wedge...\wedge e_{u_\mu},e_{v_1}\wedge...\wedge e_{v_\mu})$$ be the totally anti-symmetric tensor. Let $I$ and $J$ be $m$ tuples of indices indexing an orthonormal frame for $T(M)$ and let $A$ and $B$ be $m-1$ tuples of indices indexing an orthonormal frame for $T(\partial M)$. Set \begin{eqnarray*} &&\mathcal{R}_{J,s}^{I,t}:=  R_{i_{s}i_{s+1}j_{s+1}j_s}...R_{i_{t-1}i_tj_tj_{t-1}},\\
&&\mathcal{R}_{B,s}^{A,t}:=  R_{a_{s}a_{s+1}b_{s+1}b_s}...R_{a_{t-1}a_tb_tb_{t-1}},\\
&&\mathcal{L}_{B,s}^{A,t}:=L_{a_{s}b_{s}}...L_{a_{t}b_{t}}.\end{eqnarray*}
Since the empty product is $1$, we set
$\mathcal{R}_{J,s}^{I,t}=1$, $\mathcal{R}_{B,s}^{A,t}=1$, and
$\mathcal{L}_{B,s}^{A,t}=1$ if $t<s$.

We refer to \cite{GKVZ02} for the proof of the following result.
It establishes vanishing theorems which generalize previous
results of \cite{ABP73,PG73,PG75,Pa70} to the twisted setting. It
also identifies the local index density in the twisted setting.

\begin{theorem}\label{thm1.1}
\begin{enumerate}
\item If $n$ is odd or if $n<m$, then $a_{n,m}^{d+\delta}(\phi,g)=0$. \item If $m$ is odd, then $a_{n,m}^{d+\delta}(0,g)=0$ for any $n$. \item If $n-k<m$, then $a_{n,m,k}^{d+\delta}(\phi,g)=0$. \item $a_{2\bar m,2\bar m}^{d+\delta}(\phi,g)=\textstyle \frac1{\pi^{\bar m}8^{\bar m}\bar m!} \varepsilon_J^I\mathcal{R}_{J,1}^{I,m}$.
\item $a_{m,m,0}^{d+\delta}(\phi,g)  =\textstyle\sum_k\frac1{\pi^k8^kk!(m-1-2k)!
 \vol(S^{m-1-2k})}\varepsilon_B^A \mathcal{R}_{B,1}^{A,2k}\mathcal{L}_{B,2k+1}^{A,m-1}$.
\end{enumerate}\end{theorem}

The fact that the local index density is not dependent on the
dilaton field has important physical consequences \cite{GKVZ02}.
One can also combine Equation
(\ref{eqn1.a}) with Theorem \ref{thm1.1} to obtain a heat equation proof of the Chern-Gauss-Bonnet theorem \cite{C44,C45} for manifolds with boundary: \begin{eqnarray*} \chi(M^{2\bar m})&=&\textstyle\int_M\frac1{\pi^{\bar m}8^{\bar m}\bar m!} \varepsilon_J^I\mathcal{R}_{J,1}^{I,m}dx\\
&+&\textstyle\sum_k\int_{\partial M}\frac1{\pi^k8^kk!(2\bar
m-1-2k)!  \vol(S^{2\bar m-1-2k})}\varepsilon_B^A
\mathcal{R}_{B,1}^{A,2k}\mathcal{L}_{B,2k+1}^{A,2\bar m-1}dy,\\
\chi(M^{2\bar m+1})&=&\textstyle\sum_k\int_{\partial M}
\frac1{\pi^k8^kk!(2\bar m-2k)!\vol(S^{2\bar m-2k})}\varepsilon_B^A
\mathcal{R}_{B,1}^{A,2k}\mathcal{L}_{B,2k+1}^{A,2\bar m}dy.
\end{eqnarray*}

By Theorem \ref{thm1.1}, the first non-trivial `divergence' terms
can first arise in the supertrace when $n=m+1$. Let `;' and `:'
denote multiple covariant differentiation with respect to the
Levi-Civita connections on $M$ and on $\partial M$, respectively.
By Theorem \ref{thm1.1}, $a_{m+1,m}^{d+\delta}(\phi,g)=0$ if $m$
is even. Furthermore $a_{m+1,m,k}^{d+\delta}(\phi,g)=0$ if
$k\ge2$. The following is the main result of this paper:

\begin{theorem}\label{thm1.2}
\begin{enumerate}\item
$a_{2\bar m+2,2\bar m+1}^{d+\delta}(\phi,g)=
 \frac{1}
{\sqrt{\pi}\pi^{\bar m}8^{\bar m}\bar m!}
 \varepsilon_J^I
\phi_{;i_1j_1}\mathcal{R}_{J,2}^{I,m}$.
\item $a_{m+1,m,0}^{d+\delta} (\phi , g)=\sum_k\frac{1} {\sqrt\pi\pi^k8^kk!\vol(S^{m-2k-2})(m-2k-2)!}
\varepsilon_B^A\phi_{;a_1b_1}
\mathcal{R}_{B,2}^{A,2k+1}\mathcal{L}_{B,2k+2}^{A,m-1}$\newline$+\sum_{2k<m-3}
\frac{1}
{2\sqrt\pi\pi^k8^kk!\vol(S^{m-2k-2})(m-2k-2)!}\newline\phantom{................................}\cdot
\varepsilon_B^A\{\mathcal{R}_{B,1}^{A,2k}R_{a_{2k+1}a_{2k+2}b_{2k+2}m}\mathcal{L}_{B,2k+3}^{A,m-1}\}_{:b_{2k+1}}$,
\item $a_{m+1,m,1}^{d+\delta} (\phi , g )
=\sum_k
 \frac{\sqrt\pi}
{8^k\pi^kk!\vol(S^{m-2k})(m-2k)!}
\varepsilon_B^A\mathcal{R}_{B,1}^{A,2k}\mathcal{L}_{B,2k+1}^{A,m-1}$.\end{enumerate}\end{theorem}

Let $M$ be a closed manifold. The local index density for the
untwisted de Rham complex was identified in dimension $2$ by
McKean and Singer \cite{McSi67} and in arbitrary dimensions by
Atiyah, Bott, and Patodi \cite{ABP73}, by Gilkey \cite{PG73}, and
by Patodi \cite{Pa70}. The case of manifolds with boundary was
studied in \cite{PG75}. We also refer to \cite{BGV91,B87,Me93} for
other treatments of the local index theorem.

Patodi's approach involved a direct calculation analyzing
cancellation formulas for the fundamental solution of the heat
equation. Atiyah, Bott, and Patodi used invariance theory to
identify the local index density for the twisted signature and
twisted spin complexes. They then expressed the de Rham complex
locally in terms of the spin complex twisted by a suitable
coefficient bundle. Neither of these approaches seems particularly
well adapted to the twisted setting. In particular, since the
operator $d_\phi$ relies on the $\mathbb{Z}$ grading of the de
Rham complex, it is not described in terms of an operator on the
twisted signature or spin complexes. Thus we choose in
\cite{GKVZ02} to generalize the approach of \cite{PG73} to
determine the local index density for the twisted de Rham complex.

There are explicit combinatorial formulas
\cite{BG90,BGKV99,kbook01} for the heat trace invariants of order
$n\le 5$, see the discussion in Section \ref{Sect2} for further
details. However, these formulas become very complicated and it
seems hopeless to prove Theorem \ref{thm1.2} by an explicit
computation.

 The approach taken by Gilkey in \cite{PG73} suffered from the disadvantage that the techniques involved were rather ad hoc and cumbersome as they did not make full use of the machinery of invariance theory developed by H. Weyl \cite{We46}. In the present paper, we use both the first and second main theorems of invariance theory; this is the crucial new feature of our analysis. Let $$\begin{array}{ll} \mathcal{E}_{m+1,m}&:=\varepsilon_I^J\phi_{;i_1j_1}\mathcal{R}_{J,2}^{I,m},\vphantom{\vrule height 12pt}\\ \mathcal{F}_{m-1,m}^k&:=\varepsilon_B^A\mathcal{R}_{B,1}^{A,2k}
\mathcal{L}_{B,2k+1}^{A,m-1},\vphantom{\vrule height 12pt}\\
\mathcal{F}_{m,m}^{1,k}&:=
\varepsilon_B^A\mathcal{R}_{B,1}^{A,2k}\phi_{;a_{2k+1}b_{2k+1}}
 \mathcal{L}^{A,m-1}_{B,2k+2},\vphantom{\vrule height 12pt}\\ \mathcal{F}_{m,m}^{2,k}&:=  \varepsilon_B^A\mathcal{R}_{B,1}^{A,2k}\phi_{;a_{2k+1}}\phi_{;b_{2k+1}}
 \mathcal{L}^{A,m-1}_{B,2k+2},\vphantom{\vrule height 12pt}\\ \mathcal{F}_{m,m}^{3,k}&:=  \varepsilon_B^A\{\mathcal{R}_{B,1}^{A,2k}R_{a_{2k+1}a_{2k+2}b_{2k+2}m}
 \mathcal{L}_{B,2k+3}^{A,m-1}\}_{:b_{2k+1}}\,.\vphantom{\vrule height 12pt} \end{array}$$

\begin{lemma}\label{lem1.3}
There exist universal constants so that:
\begin{enumerate}
\item If $m$ is odd, then $a_{m+1,m}^{d+\delta} (\phi , g)=c_{m+1,m}\mathcal{E}_{m+1,m}$. \item $a_{m+1,m,1}^{d+\delta} (\phi , g) =\sum_kc_{m+1,m,1}^k\mathcal{F}_{m-1,m}^k$.
\item $a_{m+1,m,0}^{d+\delta} (\phi , g) =\sum_{i,k}c_{m+1,m,0}^{i,k}\mathcal{F}_{m,m}^{i,k}$.
\end{enumerate}\end{lemma}

This reduces the proof of Theorem \ref{thm1.2} to the evaluation
of the unknown universal coefficients. Here is a brief guide to
the remainder of the paper. In Section \ref{Sect2}, we review the
properties of the heat trace invariants which we will need. In
Section \ref{Sect3}, we use invariance theory to establish Lemma
\ref{lem1.3}. In Section \ref{Sect4}, we employ product formulas,
special case calculations, and functorial properties to derive
some technical results concerning the universal coefficients of
Lemma \ref{lem1.3}. We then combine these results to complete the
proof of Theorem \ref{thm1.2} in Section \ref{Sect5}.

\section{Formulas for the heat trace asymptotics}\label{Sect2}

Let $D$ be an arbitrary operator of Laplace type on a vector
bundle $V$. There is a canonical connection \cite{G94} $\nabla$ on
$V$ which we use to differentiate tensors of all types and a
canonical endomorphism $E$ of $V$ so that $$Du=-(u_{;ii}+Eu)\,.$$

We impose mixed boundary conditions. Let $\chi$ be an endomorphism
of $V|_{\partial M}$ so $\chi^2=1$. Decompose $\chi=\Pi_+-\Pi_-$
where $\Pi_\pm:=\frac12(\Id\pm\chi)$ are the projections on the
$\pm1$ eigenspaces of $\chi$. Let $\SS$ be an auxiliary
endomorphism of $\Pi_+$. We extend $\chi$ and $S$ to be parallel
with respect to the geodesic normal vector field $e_m$ near
$\partial M$. We impose Robin boundary conditions on
$V_+:=\operatorname{Range}(\Pi_+)$ and Dirichlet boundary
conditions on $V_-:=\operatorname{Range}(\Pi_-)$ to define the
mixed boundary operator:
$$\BB:=\{\Pi_+(\nabla_{e_m}+\SS)\oplus\Pi_-\}|_{\partial M}\,.$$

Let $\Omega_{ij}$ be the components of the curvature endomorphism
defined by $\nabla$. We refer to \cite{BG90} for the proof of the
following result which expresses the heat trace asymptotics in
terms of this formalism for $n\le3$:

\begin{lemma}\label{lem2.1}
\begin{enumerate}\item $a_0(f,D,\BB)=(4\pi)^{-m/2}\int_{M}\Tr(f\operatorname{Id})dx$.
\smallbreak\item $a_1(f,D,\BB)=
(4\pi)^{-(m-1)/2}\frac14\int_{\partial M}\Tr(f\chi)$dy.
\smallbreak\item $a_2(f,D,\BB)=(4\pi)^{-m/2}\frac16
\int_{M}\Tr\{f(6E+R_{ijji}\operatorname{Id})\}dx$\newline
$+(4\pi)^{-m/2}\frac16\int_{\partial M}
\Tr\{f(2L_{aa}\Id+12\SS)+3f_{;m}\chi\}dy$. \smallbreak\item
$a_{3}(f,D,\BB)=\textstyle(4\pi)^{-(m-1)/2}\frac1{384}\int_{\partial
M}
 \Tr\{f(96 \chi E+16 \chi R_{ijji} +8 \chi R_{amam}$\newline $+[13 \Pi_+-7 \Pi_-]L_{aa}L_{bb}+[2 \Pi_++10 \Pi_-]L_{ab}L_{ab}+96\SS L_{aa}  +192\SS^{ 2}$\newline $-12 \chi_{:a} \chi_{:a})+f_{;m}([6 \Pi_++30 \Pi_-]L_{aa}+96\SS)
+24 \chi f_{;mm} \}dy$. \end{enumerate}\end{lemma}

Similar formulas are available \cite{BG90,BGKV99,kbook01,DV94} for
$n=4,5$. What is crucial to our analysis, however, is the general
form of these expressions. They are the trace of certain
non-commutative polynomials in the covariant derivatives of the
variables $\{R,E,\Omega,\SS,L,\chi\}$ with indices contracted in
pairs.

To apply Lemma \ref{lem2.1} to the setting at hand, we must
identify the structures which are involved for the twisted
Laplacian. Let $\xxe_i:\omega\rightarrow e_i\wedge\omega$ be left
exterior multiplication by the covector $e_i$ and let $\xxi_i$ be
the dual operator, left interior multiplication by $e_i$. Let
$\gamma_i=\xxe_i-\xxi_i$ give the associated Clifford module
structure on the exterior algebra. Extend the Levi-Civita
connection to act on tensors of all types and let $\Omega_{ij}$ be
the associated curvature operator.

\begin{lemma}\label{lem2.2}
\begin{enumerate}
\item $\Delta_{\phi,g}=\Delta_g
+\phi_{;i}\phi_{;i}\cdot\Id+\phi_{;ji}(\xxe_i\xxi_j-\xxi_j\xxe_i)$.
\smallbreak\item The Levi-Civita connection is the connection
associated to $\Delta_{\phi,g}$. \smallbreak\item
$E_{\phi,g}:=-\textstyle\frac12\gamma_i\gamma_j\Omega_{ij}
-\phi_{;i}\phi_{;i}-\phi_{;ji}(\xxe_i\xxi_j-\xxi_j\xxe_i)$ is the
endomorphism for $\Delta_{\phi,g}$. \smallbreak\item Absolute
boundary conditions are defined by taking
$$\qquad\chi:=\left\{\begin{array}{l}
+1\text{ on }\Lambda(\partial M)\\
-1\text{ on }\Lambda(\partial M)^\perp
 \end{array}\right\}
\quad\text{and}\quad \SS:=\left\{\begin{array}{rl}
 -L_{ab}\xxe_b\xxi_a&\text{on }\Lambda(\partial M)\\
 0&\text{on }\Lambda(\partial M)^\perp\end{array}\right\}\,.$$ \item $\chi_{:a}=2L_{ab}(\xxe_b\xxi_m+\xxe_m\xxi_b)$.
\end{enumerate}
\end{lemma}

\medbreak\noindent{\it Proof.} The classical formula
$d+\delta_g=\xxe_i\nabla_{e_i}-\xxi_j\nabla_{e_j}$ extends to the
twisted setting:
$$d_\phi+\delta_{\phi,g}=\xxe_i\nabla_{e_i}+\xxe_i\phi_{;i}-\xxi_i\nabla_{e_i}+\xxi_i\phi_{;i}\,.$$
We use the commutation rules
$\xxe_i\xxi_j+\xxi_j\xxe_i=\delta_{ij}$, the fact that
$\nabla\xxe=0$, and the fact that $\nabla\xxi=0$ to prove
Assertion (1) by computing: \begin{eqnarray*}
\Delta_{\phi,g}&=&\Delta_g+\xxe_i\nabla_{e_i}\xxi_j\phi_{;j}+
\xxi_j\phi_{;j}\xxe_i\nabla_{e_i} -\xxi_i\nabla_{e_i}\xxe_j\phi_{;j}\\
&&-\xxe_j\phi_{;j}\xxi_i\nabla_{e_i}
+(\xxe_i\xxi_j+\xxi_j\xxe_i)\phi_{;i}\phi_{;j}\\
&=&\Delta_g+(\xxe_i\xxi_j+\xxi_j\xxe_i-\xxi_i\xxe_j-\xxe_j\xxi_i)\phi_{;j}\nabla_{e_i}
 +(\xxe_i\xxi_j-\xxi_i\xxe_j)\phi_{;ji}+\phi_{;i}\phi_{;i}\\
&=&\Delta_g+(\xxe_i\xxi_j-\xxi_i\xxe_j)\phi_{;ji}+\phi_{;i}\phi_{;i}.
\end{eqnarray*}
This shows that the associated connection does not depend on
$\phi$ and hence is the Levi-Civita connection \cite{G94}. Since
the standard Weitzenb\"ock formulas yield
$E(\Delta_g)=-\textstyle\frac12\gamma_i\gamma_j\Omega_{ij}$,
Assertion (3) follows.

We refer to \cite{BG90} for the proof of Assertion (4). Let
$\omega_+:=e^{a_1}\wedge...\wedge e^{a_\ell}$ and let
$\omega_-:=e^m\wedge\omega_+$. We then have
$\chi\omega_\pm=\pm\omega_\pm$. We use Assertion (4) to prove
Assertion (5) by computing: \begin{eqnarray*}
&&(\nabla_{e_a}\chi-\chi\nabla_{e_a})\omega_+=(\Gamma_{abc}\xxe_c\xxi_b+\Gamma_{abm}\xxe_m\xxi_b-
 \Gamma_{abc}\xxe_c\xxi_b+\Gamma_{abm}\xxe_m\xxi_b)\omega_+,\\
&&\qquad\qquad\qquad\quad\phantom{A.}=2L_{ab}\xxe_m\xxi_b\omega_+,\\
&&(\nabla_{e_a}\chi-\chi\nabla_{e_a})\omega_-
 =(-\Gamma_{abc}\xxe_c\xxi_b-\Gamma_{amb}\xxe_b\xxi_m+\Gamma_{abc}\xxe_c\xxi_b-\Gamma_{amb}\xxe_b\xxi_m)\omega_-\\
&&\qquad\qquad\qquad\quad\phantom{A.}=2L_{ab}\xxe_b\xxi_m\omega_-\,.\qquad\qquad\qedbox
\end{eqnarray*}

We now discuss functorial properties of the supertrace
asymptotics.

\begin{lemma}\label{lem2.3}
\begin{enumerate}
\smallbreak\item On the circle,
$a_{2,1}^{d+\delta}=\frac1{\sqrt\pi}\phi_{;11}$. \smallbreak\item
We have
$a_{n,m}^{d+\delta}(\phi,g)(x)=(-1)^ma_{n,m}^{d+\delta}(-\phi,g)
(x)$. \smallbreak\item We have $\int_{\partial
M}a_{m+1,m,0}^{d+\delta}(0,g)dy=0$.
\item Let $(M,\phi,g):=(M_1\times
M_2,\phi_1+\phi_2,g_1+g_2)$ where $\partial M_1=\emptyset$. Then
\begin{enumerate} \smallbreak\item
 $a_{n,m}^{d+\delta}(\phi,g)  =\textstyle\sum_{n_1+n_2=n}a_{n_1,m_1}^{d+\delta}(\phi_1,g_1)\cdot
 a_{n_2,m_2}^{d+\delta}(\phi_2,g_2)$,\phantom{aa.}
\smallbreak\item
 $a_{n,m,k}^{d+\delta}(\phi,g)  =\textstyle\sum_{n_1+n_2=n}a_{n_1,m_1}^{d+\delta}(\phi_1,g_1)\cdot
 a_{n_2,m_2,k}^{d+\delta}(\phi_2,g_2)$.
\end{enumerate}
\end{enumerate}
\end{lemma}

\begin{proof} Asssertion (1) follows from Lemma \ref{lem2.1} (3) and from Lemma \ref{lem2.2} (3).

 Since the interior invariants $a_{n,m}^{d+\delta}(\phi,g)$ are local, we may suppose without loss of generality that $M$ is a closed orientable manifold in the proof of Assertion (2). Let $\tilde\star_g$ be the normalized Hodge operator defined by the metric. Then, the normalizations having taken into account the sign conventions, the usual intertwining relations extend to the twisted context to show $$\tilde\star_g^2=\text{id},\quad\tilde\star_gd_\phi\phantom{.}\tilde\star_g=\delta_{-\phi,g},\quad\text{and}\quad
\tilde\star_g\phantom{.}\delta_{\phi,g}\phantom{.}\tilde\star_g=d_{-\phi}\,.$$
 Assertion (2) now follows from the
intertwining relationship:
$$\tilde\star_g\phantom{.}\Delta^{\phantom{..}p}_{\phi,g}\phantom{.}\tilde\star_g=
\Delta^{\phantom{.}m-p}_{\phantom{.}-\phi,g}\,.$$ We note that
$\tilde\star_g$ intertwines absolute and relative boundary
conditions; thus we can not conclude a similar equivariance
property for the boundary invariants.

We use Theorem \ref{thm1.1} to see that
$a_{m+1,m}^{d+\delta}(0,g)=0$ regardless of the parity of $m$. As
the interior invariant vanishes pointwise, the boundary integral
vanishes by Equation (\ref{eqn1.a}).

To prove Assertion (4), we decompose
$$\Lambda(M)=\Lambda(M_1)\otimes\Lambda(M_2),\
d_\phi=d_1+d_2,\text{ and } \delta_{\phi,g}=\delta_1+\delta_2$$
 where, on
$C^\infty(\Lambda^p(M_1)\otimes\Lambda^q(M_2))$, we have
$$\begin{array}{ll}
d_1:=d_{\phi_1}\otimes\Id,&d_2:=(-1)^{p\phantom{.}}\Id\otimes
d_{\phi_2},\\
\delta_1:=\delta_{\phi_1,g_1}\otimes\Id,&\delta_2:=(-1)^{p\phantom{.}}\Id\otimes\delta_{\phi_2,g_2}.
\end{array}
$$
Consequently these operators satisfy the commutation relations:
$$d_1d_2+d_2d_1=0,\ d_1\delta_2+\delta_2d_1=0,\
\delta_1d_2+d_2\delta_1=0,\
\delta_1\delta_2+\delta_2\delta_1=0\,.$$ Thus the associated
Laplacian and fundamental solution of the heat equation decompose
in the form \begin{eqnarray*}
&&\Delta_{\phi,g}^p=\oplus_{p=p_1+p_2}\Delta_{\phi_1,g_1}^{p_1}\otimes\Id+\Id\otimes
\Delta_{\phi_2,g_2}^{p_2},\\
&&e^{-t\Delta_{\phi,g,\BB_a}^p}=\oplus_{p=p_1+p_2}e^{-t\Delta_{\phi_1,g_1}^{p_1}}\otimes
 e^{-t\Delta_{\phi_2,g_2,\BB_a}^{p_2}}\,.\end{eqnarray*}
Let $f=f_1f_2$ where $f_i\in C^\infty(M_i)$. We then have
$$\Tr_{L^2}\{fe^{-t\Delta_{\phi,g,\BB_a}^p}\}=\textstyle\sum_{p=p_1+p_2}
\Tr_{L^2}\{f_1e^{-t\Delta_{\phi_1,g_1}^{p_1}}\}\cdot
\Tr_{L^2}\{f_2e^{-t\Delta_{\phi_2,g_2,\BB_a}^{p_2}}\}\,.
$$
Assertion (4) now follows by equating coefficients in the
asymptotic expansion of the supertrace. \end{proof}

\section{Invariance theory}\label{Sect3}

Let $V$ be an $m$ dimensional real vector space which is equipped
with a positive definite inner product $g(\cdot,\cdot)$. Let
$O(V)$ be the associated orthogonal group. One says that a
polynomial map $f:\times^kV\rightarrow\mathbb{R}$ is an {\it
orthogonal invariant} if $$f(\xi v^1,...,\xi
v^k)=f(v^1,...,v^k)\quad\forall\xi\in O(V)\quad\text{and}
\quad\forall(v^1,...,v^k)\in\times^kV\,.$$

Weyl's first theorem of invariants \cite{We46} (Theorem 2.9.A) is
the following:

\begin{theorem}\label{thm3.1}
 Every orthogonal invariant
depending on $k$ vectors $(v_1, ..., v_k)$ in $\times^kV$ is
expressible in terms of the $k^2$ scalar invariants $g(v_i,v_j)$.
\end{theorem}

Let $\mathcal{I}_{k,m}$ be the set of all multilinear invariant
maps from $\times^kV$ to $\mathbb{R}$; only the dimension $m$ of
$V$ is really relevant so we suppress $V$ from the notation. Given
our interest is in $O(V)$ and not $SO(V)$ invariance, we have
$\mathcal{I}_{k,m}=\{0\}$ if $k$ is odd. Consequently, we shall
suppose that $k$ is even henceforth. Let $\Sigma_k$ be the group
of all permutations of the set $\{1,...,k\}$. We define a
multi-linear invariant map $p_{k,\sigma}$ for any permutation
$\sigma\in\Sigma_k$ by setting:
$$p_{k,\sigma}(v_1,...,v_k):=g(v_{\sigma(1)},v_{\sigma(2)})\cdot\cdot\cdot
g(v_{\sigma(k-1)},v_{\sigma(k)})\,.$$

\begin{theorem}\label{thm3.2} $\mathcal{I}_{k,m}=\operatorname{span}_{\sigma\in
\Sigma_k}\{p_{k,\sigma}\}$.\end{theorem}

\begin{proof} We use Theorem \ref{thm3.1} to express $p\in\mathcal{I}_{k,m}$ in terms of monomials involving the inner products $g(v_i,v_j)$. Since $p$ is multi-linear, $$p(cv_1,v_2,...,v_k)=cp(v_1,v_2,...,v_k)\,.$$
Consequently we need only consider monomials where the variable
$v_1$ appears exactly once as otherwise we contradict
multi-linearity. A similar observation holds for the remaining
indices and these are exactly the expressions $p_{k,\sigma}$
defined above. \end{proof}

In view of Theorem \ref{thm3.2}, one says `invariant multilinear maps are given by contractions of indices' as, relative to an orthonormal basis, the inner products involved correspond to contraction of indices in pairs. Let $\{e_i\}$ be an orthonormal basis for the vector space $V$ and let $\omega=\omega_{i_1i_2...i_k}e_{i_1}\otimes...\otimes e_{i_k}\in\otimes^kV$. We have, for example: \begin{eqnarray*} &&\mathcal{I}_{2,m}:=\PSpan\{\omega\rightarrow\omega_{ii}\},\quad\text{and}\\
&&\mathcal{I}_{4,m}:=\PSpan\{\omega\rightarrow\omega_{iijj},\
\omega\rightarrow\omega_{ijij},\
\omega\rightarrow\omega_{ijji}\}\,.
\end{eqnarray*}

Let $\mathcal{P}_{n,m}$ be the space of invariant polynomials
which are homogeneous of weight $n$ in the derivatives of the
metric tensor. Atiyah, Bott, and Patodi \cite{ABP73} applied this
formalism to study these spaces. In geodesic coordinate systems,
all jets of the metric can be computed in terms of the covariant
derivatives of the curvature tensor and vice versa. Thus, for
example, if $n=4$, an invariant $P\in\mathcal{P}_{4,m}$ can be
regarded as a map from a certain subspace
$$W\subset\{\otimes^6T(M)\}\oplus\{\otimes^8T(M)\}$$
to $\mathbb{R}$ which is invariant under the action of the orthogonal group; here $W$ is generated by the algebraic covariant derivatives $\nabla^2R\subset\otimes^6T(M)$ and by the algebraic curvature tensors $R\otimes R\subset\otimes^8T(M)$. As the subspace $W$ is orthogonally invariant, extending $P$ to be zero on $W^\perp$ defines an orthogonally invariant map to which Theorem \ref{thm3.2} applies. Thus, for example, after taking into account the appropriate curvature symmetries, one has: \begin{eqnarray*} &&\mathcal{P}_{2,m}=\PSpan\{\tau:=R_{ijji}\},\\
&&\mathcal{P}_{4,m}=\PSpan\{\tau^2,\ |\rho^2|:=R_{ijjk}R_{illk},\
|R|^2:=R_{ijkl}R_{ijkl}, \ \Delta\tau:=-R_{ijji;kk}\}.
\end{eqnarray*} This analysis extends to form valued invariants with coefficients in an auxiliary vector bundle and gives rise to a heat equation proof of the index theorem for the classical elliptic complexes \cite{ABP73}.

What is relevant to our analysis, however, is Weyl's second main
theorem \cite{We46} (Theorem 2.17.A).

\begin{theorem}\label{thm3.3}
Every relation among scalar products
is an algebraic consequence of the relation $$0=\det\left(\begin{array}{llll} g(v_1,w_1)&g(v_2,w_1)&...&g(v_{m+1},w_1)\\
g(v_1,w_2)&g(v_2,w_2)&...&g(v_{m+1},w_2)\\
...&...&...&...\\
g(v_1,w_{m+1})&g(v_2,w_{m+1})&...&g(v_{m+1},w_{m+1})
\end{array}\right).$$\end{theorem}

We remark that this relation can also be expressed in the form:
\begin{equation}\label{eqn3.a} 0=g(v_1\wedge...\wedge
v_{m+1},w_1\wedge...\wedge w_{m+1}). \end{equation}

Let $W$ be a vector space of dimension $m-1$. Choose an inner
product preserving inclusion $i:W\subset V$ which embeds
$O(W)\subset O(V)$. We define the restriction map
$$r:\mathcal{I}_{k,m}\rightarrow\mathcal{I}_{k,m-1}$$ which is
characterized dually by the property:
$$r(p)(w_1,...,w_k)=p(i(w_1),...,i(w_k))\,.$$ If $p$ is given by
contractions of indices which range from $1$ to $m$, then $r(p)$
is given by restricting the range of summation to range from $1$
to $m-1$. Consequently, the map $r$ is surjective. If $k\ge2m$ and
if $\sigma\in \Sigma_k$, define:
\begin{eqnarray*}\Theta_{k,m,\sigma}(v_1,...,v_k):&=&g(v_{\sigma(1)}\wedge...\wedge
v_{\sigma(m)}, v_{\sigma(m+1)}\wedge...\wedge v_{\sigma(2m)})\\
&\times&g(v_{\sigma(2m+1)},v_{\sigma(2m+2)})\cdot\cdot\cdot
g(v_{\sigma(k-1)},v_{\sigma(k)}). \end{eqnarray*}

\begin{theorem}\label{thm3.4}
Let $m\ge2$.\begin{enumerate}
\item $r:\mathcal{I}_{k,m}\rightarrow\mathcal{I}_{k,m-1}$ is surjective. \item $r:\mathcal{I}_{k,m}\rightarrow\mathcal{I}_{k,m-1}$ is injective if $k<2m$. \item If $k\ge 2m$, then $\ker(r)\cap\mathcal{I}_{k,m}=\operatorname{span}
 _{\sigma\in \Sigma_k}\{\Theta_{k,m,\sigma}\}$.
\end{enumerate}
\end{theorem}

\begin{proof} We have already verified Assertion (1). To prove Assertion (2), we use Theorem \ref{thm3.2} to express $p\in\mathcal{I}_{k,m}$ in terms of inner products. We use Theorem \ref{thm3.3}, after making an appropriate dimension shift, to see that $r(p)$ vanishes if and only if it can be written as sums of terms each of which is divisible by an appropriate determinant $J$ of size $m\times m$. The desired result now follows from equation (\ref{eqn3.a}) and from the same arguments used to prove Theorem \ref{thm3.2}. \end{proof}

Previously we have considered invariants of the metric alone. The
analysis extends easily to the twisted setting. We define
$$\weight(\nabla^k\phi)=k\quad\text{and}\quad\weight(\nabla^kR)=2+k.$$
Let $\mathcal{Q}_{n,m}$ be the space of all $O(m)$ invariant polynomials of total weight $n$ in the components of $R$, the covariant derivatives of $R$, and the covariant derivatives of $\phi$. We do not admit $\phi$ as a variable. Furthermore we require that each monomial either does not involve the covariant derivatives of $\phi$ at all or involves at least two covariant derivatives of $\phi$. We use the $\mathbb{Z}_2$ action $\phi\rightarrow -\phi$ to decompose \begin{eqnarray*} &&\mathcal{Q}_{n,m}=\mathcal{Q}_{n,m}^+\oplus\mathcal{Q}_{n,m}^-\quad\text{where}\\
&&\mathcal{Q}_{n,m}^\pm:=\{Q\in\mathcal{Q}_{n,m}:Q(\phi,g)=\pm
Q(-\phi,g)\}\,. \end{eqnarray*} The restriction map in Theorem
\ref{thm3.4} induces natural surjective maps
$$r:\mathcal{Q}_{n,m}^\pm\rightarrow\mathcal{Q}_{n,m-1}^\pm\rightarrow0\,.$$

If $(N,\phi_N,g_N)$ are structures in dimension $m-1$, then we can
define corresponding structures in dimension $m$ by setting
$$(M,\phi_M,g_M):=(N\times S^1,\phi_N,g_N+d\theta^2)\,.$$ If
$y\in\partial N$ is the point of evaluation, let $(y,1)\in\partial
M$ be the corresponding point of evaluation -- it does not matter
which point is chosen on the circle owing to the rotational
symmetry. The restriction map
$r:\mathcal{Q}_{n,m}\rightarrow\mathcal{Q}_{n,m-1}$ is then
characterized dually by the formula:
$$r(Q)(\phi_N,g_N)(y)=Q(\phi_N,g_N+d\theta^2)(y,1)\,.$$

\begin{lemma}\label{lem3.5}
\begin{enumerate}
\item If $m$ is even, then $a_{n,m}^{d+\delta}(\phi,g)\in\mathcal{Q}_{n,m}^+\cap\ker r$. \item If $m$ is odd, then $a_{n,m}^{d+\delta}(\phi,g)\in\mathcal{Q}_{n,m}^-\cap\ker r$. \end{enumerate}\end{lemma}

\begin{proof} Standard arguments \cite{G94} show the invariants $a_{n,m}^{d+\delta}(\phi,g)$ are homogeneous of weight $n$ in the jets of the metric and of $\phi$. Let $\nabla$ be the Levi-Civita connection on $\Lambda M$. By Lemma \ref{lem2.2} (1), $$\Delta_{\phi,g}^p=\Delta_g+\ffrac12\gamma_i\gamma_j\Omega_{ij}
+\phi_{;i}\phi_{;i}-\phi_{;ji}(\xxe_i\xxi_j-\xxi_j\xxe_i)\,.
$$
Thus the undifferentiated variable $\phi$ does not play a role in
these invariants. Furthermore, either at least $2$ covariant
derivatives of $\phi$ appear or only the curvature $R$ appears in
each Weyl monomial of $a_{n,m}^{d+\delta}(\phi,g)$. This shows
that $$a_{n,m}^{d+\delta}(\phi,g)\in\mathcal{Q}_{n,m}\,.$$

We use Lemma \ref{lem2.3} (2) to see that
$a_{n,m}^{d+\delta}(\phi,g)$ is an odd function of $\phi$ if $m$
is odd and an even function of $\phi$ if $m$ is even. To complete
the proof, we must show $ra_{n,m}^{d+\delta}=0$. Suppose that
$M=N\times S^1$ has the product metric and that $\phi=\phi_N$ is
independent of the angular parameter $\theta\in S^1$. As
$\phi_{S^1}=0$, we use Lemma \ref{lem2.3} (3) to see
$a_{n,1}^{d+\delta}(0,g_{S^1})=(-1)^1a_{n,1}^{d+\delta}(0,g_{S^1})=0$
for all $n$. Thus Lemma \ref{lem2.3} (4a) implies that
$a_{n,m}^{d+\delta}(\phi_M,g_N)=0$. This shows that
$ra_{n,m}^{d+\delta}=0$. \end{proof}

Assertion (1) of Lemma \ref{lem1.3} will follow from the following
result: \begin{lemma}\label{lem3.6} If $m$ is odd, then
$\mathcal{Q}_{m+1,m}^-\cap\ker r=\PSpan\{\mathcal{E}_{m+1,m}\}$.
\end{lemma}

\begin{proof} Let $0\ne Q\in\mathcal{Q}_{m+1,m}^-$. Let $A$ be a monomial of $Q$ of the form $$ A=\phi_{;\alpha_1}...\phi_{;\alpha_u}R_{i_1j_1k_1\ell_1;\beta_1}...R_{i_vj_vk_v\ell_v;\beta_v}
$$
where $\alpha_\mu$ and $\beta_\nu$ denote appropriate collections
of indices. Then $$
m+1=\weight(A)=\textstyle\sum_\mu|\alpha_\mu|+\sum_\nu(2+|\beta_\nu|)\,.
$$
By definition, the empty sum is $0$. Thus $\sum_\mu$ is to be
ignored if $u=0$ and $\sum_\nu$ is to be ignored if $v=0$. Let $k$
be total number of indices present in $A$;
$$k:=\textstyle\sum_\mu|\alpha_\mu|+\textstyle\sum_\nu(4+|\beta_\nu|)=\weight(A)+2v=m+1+2v\,.$$

We apply Weyl's second main theorem of invariance theory as
discussed above. To ensure that $rQ=0$, we must contract $2m$
indices in $A$ using the $\varepsilon$ tensor and then contract
the remaining indices of $A$ in pairs. Consequently, at least $2m$
indices must appear in $A$ so \begin{equation}\label{eqn3.b} 2m\le
k=m+1+2v=2m+2-\textstyle\sum_\mu|\alpha_\mu|-\textstyle\sum_\nu |
\beta_\nu|\le 2m+2\,. \end{equation} Since $m$ is odd, $2m$,
$m+1+2v$, and $2m+2$ are all even. Thus only one of the two
inequalities given in Display (\ref{eqn3.b}) can be strict. As
$Q(-\phi,g)=-Q(\phi,g)$, $u$ must be {\it odd}. Thus
$$\textstyle\sum_\mu|\alpha_\mu|>0$$
 so the second inequality in Equation (\ref{eqn3.b}) is strict. Thus exactly $2m=k$ indices appear in $A$ and all are contracted using the $\varepsilon$ tensor. The first and second Bianchi identity show $R_{****;\beta}=0$ if 3 indices are alternated. Thus at most two $i$ indices and at most two $j$ indices can appear in each $R_{****;\beta}$ variable. This shows that $$|\beta_\nu|=0\quad\text{for all}\quad\nu\,.$$ Furthermore, the two possibilities are $R_{i_1i_2j_2j_1}$ or $R_{i_1j_1i_2j_2}$. The first Bianchi identity can then be used to express the second variable in terms of the first. Since $$u\le|\alpha_1|+...+|\alpha_u|=2$$
and $u$ is odd, $u=1$ and $|\alpha_1|=2$, since either 0 or at
least $2$ covariant derivatives of $\phi$ appear in each monomial
of $Q\in\mathcal{Q}_{n,m}$. Thus we are in fact dealing with a
multiple of
$\mathcal{E}_{m+1,m}:=\varepsilon_J^I\phi_{;i_1j_1}\mathcal{R}_{J,2}^{I,m}$.
\end{proof}

To complete the proof of Lemma \ref{lem1.3}, we study the boundary
invariants. Let $\tilde\nabla$ denote the Levi-Civita connection
of the boundary. We consider polynomials in the components of the
tensors $$\{R,\ \nabla R,\ \nabla^2R,\ ...\ ,\ L,\ \tilde\nabla
L,\ \tilde\nabla^2L,\ ...\ ,\ \nabla\phi,\ \nabla^2\phi,...\}\,.$$
Again, we do not introduce the variable $\phi$. We let
$$\weight(\nabla^kR):=2+k,\ \weight(\tilde\nabla^kL):=1+k,\text{
and }\weight(\nabla^k\phi)=k\,.$$ Let $\QQ_{n,m}$ be the space of
all $O(m-1)$ invariant polynomials of total weight $n$ where we
admit monomials which either do not involve the covariant
derivatives of $\phi$ at all or which involve at least two
covariant derivatives of $\phi$.

Let $\PP_{n,m}\subset\QQ_{n,m}$ be the subspace of invariants
which do not involve the covariant derivatives of $\phi$. Setting
$\phi=0$ defines a natural map from $\QQ_{n,m} $ to $\PP_{n,m} $.
If $P\in\PP_{n,m} $, then the evaluation $\xxI(P)(g)\in\mathbb{R}$
is defined by setting $$\xxI P(g):=\textstyle\int_{\partial
M}P(g)(y)dy\,.$$

By Lemma \ref{lem2.3} (3), $\xxI a_{m+1,m,0}^{d+\delta}(0,g)=0$.
The same argument as that given to establish Lemma \ref{lem3.5}
can be used to show that
$a_{n,m,k}^{d+\delta}\in\tilde{\mathcal{Q}}_{n-k-1}\cap\ker r$.
The remaining assertions of Lemma \ref{lem1.3} will now follow
from the following result:

\begin{lemma}\label{lem3.7}
\begin{enumerate}
\item $\QQ_{n,m} \cap\ker r=\{0\}$ if $n<m-1$.
\item $\QQ_{m-1,m} \cap\ker r=\PSpan_k\{\mathcal{F}_{m-1,m}^k\}$.
\item
$\QQ_{m,m} \cap\ker
r=\PSpan_k\{\mathcal{F}_{m,m}^{1,k},\mathcal{F}_{m,m}^{2,k}\}
+\{\PP_{m,m}\cap\ker r\}$. \smallbreak\item $\PP_{m,m} \cap\ker r
\cap\ker\xxI=\PSpan_k\{\mathcal{F}_{m,m}^{3,k}\}$.
\end{enumerate}
\end{lemma}

\begin{proof} Let $0\ne Q\in\QQ_{n,m} \cap\ker r$ and let $A$ be a monomial of $Q$ of weight $n$ where: \begin{eqnarray*} &&A:=\phi_{;\alpha_1}\cdot\cdot\cdot \phi_{;\alpha_u}R_{i_1j_1k_1\ell_1;\beta_1}\cdot\cdot\cdot
R_{i_vj_vk_v\ell_v;\beta_v}L_{a_1b_1:\gamma_1}
\cdot\cdot\cdot L_{a_wb_w:\gamma_w},\\
&&\textstyle
n:=\sum_\mu|\alpha_\mu|+\sum_\nu(|\beta_\nu|+2)+\sum_\sigma(|\gamma_\sigma|+1)\,.
\end{eqnarray*}

To ensure that $rQ=0$, we contract $2(m-1)$ tangential indices in
$A$ using the $\varepsilon$ tensor; the remaining tangential
indices must be contracted in pairs. Since the structure group is
$O(m-1)$, the normal index `$m$' can stand alone and unchanged.
Let $k_T$ be the total number of tangential indices in $A$, and
let $k_m$ be the total number of times the normal index $m$
appears in $A$. We estimate: \begin{eqnarray} &&2m-2\le k_T\le
k_T+k_m\nonumber  \\
&&=\textstyle\sum_\mu|\alpha_\mu|+\sum_\nu(|\beta_\nu|+4)+\sum_\sigma
(|\gamma_\sigma|+2)=n+2v+w\label{eqn3.c}
\\ &&\textstyle=2n-\sum_\mu|\alpha_\mu|-\sum_\nu|\beta_\nu|-\sum_\sigma|\gamma_\sigma|
\le2n.\nonumber
\end{eqnarray}
Assertion (1) of the Lemma follows as this is not possible if
$n<m-1$.

We set $n=m-1$ to prove Assertion (2). All the inequalities of
Display (\ref{eqn3.c}) must have been equalities so there are no
covariant derivatives and thus the $\phi$ variables do not appear.
All the indices are tangential and are contracted using the
$\varepsilon$ tensor. After using the first Bianchi identity, we
see that this leads to the invariants $\mathcal{F}_{m-1,m}^k$
which proves Assertion (2).

Let $n=m$. Display (\ref{eqn3.c}) involves a total increase of
$2$. Thus at most $2$ explicit covariant derivatives are present.
However, unless at least $2$ covariant derivatives are present,
$\phi$ is not involved and this leads to invariants in
$\PP_{m,m}\cap\ker r$. Thus we may suppose exactly $2$ explicit
covariant derivatives are present -- and all of them appear on
$\phi$. Consequently $$k_T=2m-2,\ k_m=0,\
\textstyle\sum_\mu|\alpha_\mu|=2,\ \sum_\nu|\beta_\nu|=0,\
\text{and}\ \sum_\sigma|\gamma_\sigma|=0\,.$$ Since every index is
tangential and all are contracted using the tensor $\varepsilon$,
after applying the Bianchi identities, we obtain the invariants
$\mathcal{F}_{m,m}^{1,k}$ and $\mathcal{F}_{m,m}^{2,k}$. This
completes the proof of Assertion (3).

To prove Assertion (4), we set $\phi=0$ and consider only metric
invariants. Let $\PP_{n,m}^p$ be the space of $p$ form valued
invariants which are homogeneous of degree $n$ in the derivatives
of the metric; $\PP_{n,m}=\PP_{n,m}^0$.

Let $\tilde\delta :\PP_{n,m}^p\rightarrow\PP_{n+1,m}^{p-1}$ be the
coderivative of the boundary. Results of \cite{PG75} describe the
cohomology groups of this complex. When combined with standard
methods of invariance theory they yield the following
observations:
\begin{enumerate}\item $r$ is a surjective map from $\PP_{n,m}^p$ to $\PP_{n,m-1}^p$ with $r\tilde\delta=\tilde\delta r$. \item If $n\ne m-1$, then $\PP_{n,m}^0\cap\ker\xxI=\tilde\delta\PP_{n-1,m}^1$.
\item If $n\ne m-1$, then $\PP_{n-1,m}^1\cap\ker\tilde\delta=\tilde\delta\PP_{n-2,m}^2$.
\end{enumerate}

Let $P_{m,m}\in\PP_{m,m}\cap\ker r\cap\ker\xxI $. Choose
$P_{m-1,m}^1\in\PP_{m-1,m}^1$ so $\tilde\delta
P_{m-1,m}^1=P_{m,m}$. Unfortunately, $rP_{m-1,m}^1$ need not
vanish and we must adjust $P_{m-1,m}^1$. Since $$\tilde\delta
rP_{m-1,m}^1=r\tilde\delta P_{m-1,m}^1=rP_{m,m}=0\,,$$ we may
choose $P_{m-2,m-1}^2\in\PP_{m-2,m-1}^2$ so $\tilde\delta
P_{m-2,m-1}^2=rP_{m-1,m}^1$. Since $r$ is surjective, we may
choose $P_{m-2,m}^2\in\PP_{m-2,m}^2$ so
$rP_{m-2,m}^2=P_{m-2,m-1}^2$. Then: \begin{eqnarray*}
&&\tilde\delta\{P_{m-1,m}^1-\tilde\delta
P_{m-2,m}^2\}=\tilde\delta P_{m-1,m}^1=P_{m,m},\\
&&r\{P_{m-1,m}^1-\tilde\delta
P_{m-2,m}^2\}=rP_{m-1,m}^1-\tilde\delta rP_{m-2,m}^2\\
&&\qquad=rP_{m-1,m}^1-\tilde\delta P_{m-2,m-1}^2=0\,.
\end{eqnarray*} Consequently \begin{equation}\label{eqn3.d}
\PP_{m,m}\cap\ker
r\cap\ker\xxI=\tilde\delta\{\PP_{m-1,m}^1\cap\ker r\}\,.
\end{equation}

Let $0\ne P_{m-1,m}^1\in\PP_{m-1,m}^1\cap\ker r$ and let
$$A=R_{i_1j_1k_1\ell_1;\beta_1}...R_{i_vj_vk_v\ell_v;\beta_v}L_{a_1b_1:\gamma_1}...L_{a_wb_w:\gamma_w}e^c$$
be a monomial of $P_{m-1,m}^1$. Since $rP_{m-1,m}^1=0$, we must
contract $2(m-1)$ indices in $A$ using the $\varepsilon$ tensor
and contract the remaining indices in pairs. We estimate
\begin{equation}\label{eqn3.e}\begin{array}{l}
\textstyle 2(m-1)\le k_T\le k_T+k_m=\sum_\nu(|\beta_\nu|+4)+\sum_\sigma(|\gamma_\sigma|+2)+1\\
\textstyle=m-1+2v+w+1=2(m-1)-\sum_\nu|\beta_\nu|-\sum_\sigma|\gamma_\sigma|+1
\vphantom{\vrule height 12pt}\\
\le 2(m-1)+1\,. \vphantom{\vrule height
12pt}\end{array}\end{equation} This sequence of inequalities
represents a total increase of $1$. Thus $k_T=2(m-1)$ and every
tangential index is contracted using the $\varepsilon$ tensor. We
have \begin{equation}\label{eqn3.f}
L_{c_2c_3:c_1}-L_{c_1c_3:c_2}=R_{c_1c_2c_3m}.
\end{equation}
We may therefore assume $|\gamma_\sigma|=0$ so there are no
tangential derivatives of $L$ present. If $k_m=0$, then every
index is contracted using the $\varepsilon$ tensor. Thus the
Bianchi identities show $|\beta_\nu|=0$ for all $\nu$. This means
that every inequality in Display (\ref{eqn3.e}) is an equality
which is impossible. Consequently $k_m=1$ and
$\sum_\nu|\beta_\nu|=0$. This leads to the invariants $$
\mathcal{G}_{m-1,m}^k:=\varepsilon_A^B\mathcal{R}_{B,1}^{A,2k}
R_{a_{2k+1}a_{2k+2}mb_{2k+1}}\mathcal{L}_{B,2k+3}^{A,m-1}e^{b_{2k+2}}\,.
$$
Assertion (4) now follows from Equation (\ref{eqn3.d}) since
$\tilde\delta\mathcal{G}_{m-1,m}^k=-\mathcal{F}_{m,m}^{3,k}$.
\end{proof}

\section{Product formulas, special case computations, and functorial properties}\label{Sect4} Throughout this section, we adopt the notation of Lemma \ref{lem1.3}. We begin with a result which is based on product formulas.

\begin{lemma}\label{lem4.1}
\begin{enumerate}\item If $m=2\bar m+1$, then $c_{m+1,m}=\ffrac1{\sqrt\pi8^{\bar m}\pi^{\bar m}\bar m!}$ \item If $k>0$, then $c_{m+1,m,1}^k=\frac1{\pi^k8^kk!}c_{m-2k+1,m-2k,1}^0$.
\item If $k>0$, then $c_{m+1,m,0}^{i,k}=\frac1{\pi^k8^kk!}c_{m-2k+1,m-2k,0}^{i,0}$.
\item We have $c_{m+1,m,0}^{1,0}=\frac1{\sqrt\pi}\frac1{(m-2)!\vol(S^{m-2})}$ and $c_{m+1,m,0}^{2,0}=0$. \end{enumerate}\end{lemma}

\begin{proof} Give $S^m$ and $D^m$ the standard metrics $g_{S,m}$ and $g_{D,m}$. We then have \begin{equation}\label{eqn4.zz} \varepsilon_J^I\mathcal{R}_{J,1}^{I,2\bar m}(g_{S,m})=2^{\bar m}(2\bar m)!\quad\text{and}\quad \varepsilon_A^B\mathcal{L}_{A,1}^{B,m-1}(g_{D,m})=(m-1)!\,.
\end{equation}
Let $m=2\bar m+1$. Give $M:=S^1\times S^{2\bar m}$ the product
structures where $\phi_2=0$. By Theorem \ref{thm1.1} (1) and Lemma
\ref{lem2.3} (4) we have
$$a_{m+1,m}^{d+\delta}(\phi_M,g_M)=a_{2,1}^{d+\delta}(\phi_1,g_{S,1})a_{2\bar
m,2\bar m}(0,g_{S,2\bar m})\,.$$ Consequently, by Equation
(\ref{eqn4.zz}) and by Theorem \ref{thm1.1} (4), \begin{eqnarray*}
&&a_{m+1,m}^{d+\delta}(\phi,g)=c_{m+1,m}\phi_{;11}2^{\bar m}(2\bar
m)!\\ &=&a_{2,1}^{d+\delta}(\phi_1,g_{S,1})\cdot a_{2\bar m,2\bar
m}^{d+\delta}(0,g_{S,2\bar m})
=\ffrac1{\sqrt{\pi}}\phi_{;11}\ffrac1{8^{\bar m}\pi^{\bar m}\bar
m!}2^{\bar m}(2\bar m)!\,. \end{eqnarray*} We complete the proof
of Asssertion (1) by using this relation to solve for $c_{m+1,m}$:
$$c_{m+1,m}=\ffrac1{\sqrt\pi8^{\bar m}\pi^{\bar m}\bar m!}\,.$$

Fix $k>0$. Give $M=S^{2k}\times D^{m-2k}$ the product structures
where $\phi_1=0$. We argue as in the proof of Assertion (1) to see
that:
\begin{eqnarray*} a_{m+1,m,1}^{d+\delta}(\phi_M,g_M)&=&\textstyle\sum_{j}c_{m+1,m,1}^{j}\mathcal{F}_{m,m}^j(\phi_M,g_M)\\
&=&
c_{m+1,m,1}^k2^k(2k)!\cdot(m-2k-1)!\\
&=&a_{2k,2k}^{d+\delta}(0,g_{S,2k})\cdot a_{m-2k+1,m-2k,1}^{d+\delta}(0,g_{D,m-2k})\\
&=&\ffrac1{\pi^k8^kk!}2^k(2k)!c_{2k+1,2k,1}^0(m-2k-1)!\,.
\end{eqnarray*}
This equation relates $c_{m+1,m,1}^k$ and $c_{m+1,m,1}^0$ and
thereby establishes Assertion (2); the proof of Assertion (3) is
similar.

Let $M:=S^1\times D^{m-1}$ where $\phi=\phi(\theta)$ depends only
on $S^1$. We use Theorem \ref{thm1.1} to determine
$a_{m-1,m-1,0}^{d+\delta}(0,g_{D,m-1})$. As
$a_{2,1}^{d+\delta}=\textstyle\frac1{\sqrt\pi}\phi_{;11}$, we
argue as above to see \begin{eqnarray*}
&&a_{m+1,m,0}^{d+\delta}(\phi,g)  =\{c_{m+1,m,0}^{1,0}\phi_{;11}+c_{m+1,m,0}^{2,0}\phi_{;1}\phi_{;1}\}(m-2)!\\
&=&a_{2,1}^{d+\delta}(\phi,d\theta^2) \cdot
a_{m-1,m-1,0}^{d+\delta}(0,g_{D^{m-1}})
 =\ffrac1{\sqrt\pi}\phi_{;11}\ffrac{(m-2)!}{\vol(S^{m-2})(m-2)!}\,.
\end{eqnarray*}
We solve for $c_{m+1,m,0}^{1,0}$ and $c_{m+1,m,0}^{2,0}$ to
establish Assertion (4). \end{proof}

By Lemma \ref{lem4.1}, we need only determine $c_{m+1,m,1}^0$ and
$c_{m+1,m,0}^{3,0}$ to complete the proof of Theorem \ref{thm1.2}.
As these terms do not involve $\phi$, we set $\phi=0$ henceforth.
We introduce universal constants $\cc_{n,m,k}^\nu$ so that if
$\BB$ defines mixed boundary conditions for an operator of Laplace
type, then the heat trace asymptotics have the form:
\begin{eqnarray*}
a_{n,m,k}(y,D,\BB)&=&\cc_{n,m,k}^0\Tr\{S^{n-k-1}\}+\cc_{n,m,k}^3\Tr\{E_{;m}S^{n-k-4}\}+\ldots\,.\nonumber
\end{eqnarray*}

We will use the method of universal examples to show that only
$\Tr\{S^{m-1}\}$ is relevant in computing
$\{a_{m,m,0}^{d+\delta}(0,g),a_{m+1,m,1}^{d+\delta}(0,g)\}$ and
that only $\Tr\{E_{;m}S^{m-3}\}$ is relevant in computing
$a_{m+1,m,0}^{d+\delta}(0,g)$. This will enable us to show:

\begin{lemma}\label{lem4.2}
\ \begin{enumerate}
\item If $m\ge2$, then $c_{m+1,m,1}^0=\cc_{m+1,m,1}^0$ and $\cc_{m,m,0}^0=\frac1{(m-1)!\operatorname{vol}(S^{m-1})}$.
\item $c_{4,3,0}^{3,0}=0$. If $m>3$, then $c_{m+1,m,0}^{3,0}=\cc_{m+1,m,0}^3$.
\end{enumerate}\end{lemma}

\par\noindent{\bf Remark:} The constants $\cc_{n,m,k}^0$ and $\cc_{n,m,k}^3$ have been determined in \cite{BFSV02}; after a bit of work converting from $\Gamma$ functions into volumes of spheres one checks the value of $\cc_{m+1,m,1}^0$ given here is consistent with the value given in \cite{BFSV02}; this provides a valuable check on our methodology.

\begin{proof} We shall prove Lemma \ref{lem4.2} by making a special case calculation. Let $m\ge2$. To simplify the notation, let $$P_m(g):=a_{m+1,m,1}^{d+\delta}(0,g),\quad c_m:=c_{m+1,m,1}^0, \quad\cc_m:=\cc_{m+1,m,1}^0\,.$$

Let $(y_1,...,y_{m-1})$ be the usual coordinates on
$\mathbb{R}^{m-1}$. Let $f(y)$ be a smooth even function function
of $y$ and let $$M_m:=\{(y,r)\in\mathbb{R}^m:r\ge f(y)\}\,.$$ Let
$\{A_1,...,A_{m-1}\}$ be distinct real constants. We choose $f$ so
that \begin{equation}\label{eqn4.a} f(0)=0,\quad
(\partial_i^yf)(0)=0,\quad\text{and}\quad(\partial_i^y\partial_j^yf)(0)=A_i\delta_{ij}\,.
\end{equation}
Give $\mathbb{R}^m$ the usual flat metric. Then
$L_{ij}(0)=-A_i\delta_{ij}$. We use Lemma \ref{lem1.3} to compute:
\begin{equation}\label{eqn4.b}
P_m(g)(0)=(m-1)!c_m\mathcal{A}\quad\text{where}\quad\mathcal{A}:=(-1)^{m-1}A_1...A_{m-1}\,.
\end{equation}

Because $R=0$, we have $E=0$ and $\Omega=0$. Thus there exists a
polynomial $Q_m$ of total weight $m-1$ in the tangential covariant
derivatives of $\{\chi,L,S\}$ so that
$$P_m=\textstyle\textstyle\sum_p(-1)^p\Tr_{\Lambda^p(\mathbb{R}^m)}\{Q_m(\cdot)\}\,.$$

We must control $\tilde\nabla^kL$ for $k\ge1$. Since the curvature
of $\mathbb{R}^m$ vanishes, Equation (\ref{eqn3.f}) shows that
$\tilde\nabla L$ is a totally symmetric tensor field. Since $f$ is
an even function, $\tilde\nabla^kL$ vanishes at the origin if $k$
is odd. For $k$ even, the components of $\tilde\nabla^kL(0)$ are
polynomials in the derivatives of the defining function $f$. Let
$\KK$ denote the ideal in the algebra of all polynomials in the
jets of $f$ which is generated by the monomials
$\{A_1^2,...,A_{m-1}^2\}$. In light of Equation (\ref{eqn4.b}), we
shall work modulo $\KK$ since such elements can not contribute to
$\mathcal{A}$.

We first study $\tilde\nabla^2L$. This is not a symmetric tensor field. Let $\tilde R$ be the curvature of the Levi-Civita connection of $\partial M$. Let $\{e_1,...,e_{m-1}\}$ be an orthonormal frame for the tangent bundle of the boundary so that $e_i(0)=\partial_i^y$. Then: \begin{eqnarray*} &&\tilde R_{b_1b_2b_3b_4}=L_{b_1b_4}L_{b_2b_3}-L_{b_1b_3}L_{b_2b_4},\quad\text{and}\\
&&L_{a_1a_2:a_3a_4}-L_{a_1a_2:a_4a_3}=\tilde
R_{a_3a_4a_1a_5}L_{a_5a_2}+\tilde R_{a_3a_4a_2a_5}L_{a_5a_1}\,.
\end{eqnarray*} This shows that $A_{a_5}^2$ divides $\{\tilde
R_{a_3a_4a_1a_5}L_{a_5a_2}+\tilde
R_{a_3a_4a_2a_5}L_{a_5a_1}\}(0)$. Consequently
$\tilde\nabla^2L(0)$ is totally symmetric modulo the ideal $\KK$.
Since the components of $\tilde\nabla^2L$ are linear in the $4$
jets of $f$ and quadratic in the $2$ jets of $f$, we may choose
the $4$ jets of $f$ to kill the symmetrization of
$(\tilde\nabla^2L)(0)$ and thereby ensure
$(\tilde\nabla^2L)(0)\in\KK$. Similarly, by choosing
$\tilde\nabla^{k+2}f(0)$ appropriately, we may suppose that
$$(\tilde\nabla^kL)(0)\in\KK\quad\text{for}\quad k>0\,.$$

We therefore supress $\tilde\nabla^kL$ henceforth in the proof of
Assertions (1) and (2). By Lemma \ref{lem2.2} (5),
$\chi_{:a}=2L_{ab}(\xxe_b\xxi_m+\xxe_m\xxi_b)$. Thus further
covariant differentiation of $\chi$ only involves covariantly
differentiating $\xxe_b\xxi_m+\xxe_m\xxi_b$. Thus inductively
there exist suitably chosen endomorphisms $\mathcal{E}_{\star}$ of
weight $0$ so: \begin{equation}\label{eqn4.d}
\chi_{:a_1...a_k}=L_{a_1b_1}L_{a_2b_2}...L_{a_kb_k}
\mathcal{E}_{b_1...b_k}\,.
\end{equation}

If a $\chi_{:a_1...}$ term appears, we must contract it with
another index $a_1$; Equation (\ref{eqn4.d}) contains no
$L_{a_1a_1}$ term. Consequently this contraction involves a
different variable which produces an $A_{a_1}^2$ term; such terms
can be ignored in light of Equation (\ref{eqn4.b}). Similarly
since $$S=-L_{ab}\xxe_b\xxi_a\text{ on
}\Lambda(\mathbb{R}^{m-1})\quad\text{and}\quad S=0\text{ on
}\Lambda(\mathbb{R}^{m-1})\wedge dr,
$$
$\tilde\nabla^kS$ plays no role if $k\ge1$. If a $L_{a_1b_1}$ term
appears where $a_1$ is not to be contracted with $b_1$, then $A$
must be divisible by $A_{a_1}^2$. If the term $L_{aa}$ appears in
a monomial $Q$, then we may factor $Q=L_{aa}Q_0$ and then apply
Lemma \ref{lem3.7} (1) to see the supertrace of $Q_0$ vanishes.
Thus $L$ does not appear as a variable. This shows that only the
monomial $S^{m-1}$ is relevant. Consequently $$
P_m(g)(0)=\cc_m\textstyle\textstyle\sum_p(-1)^p\Tr_{\Lambda^p(\mathbb{R}^m)}\{S^{m-1}\}(0)\,.
$$
Since $S$ is zero on $\Lambda^p(\mathbb{R}^{m-1})\wedge dr$,
\begin{equation}\label{eqn4.e}
P_m(g)(0)=\cc_m\textstyle\textstyle\sum_p(-1)^p\Tr_{\Lambda^p(\mathbb{R}^{m-1})}\{S^{m-1}\}(0)\,.
\end{equation}
We may decompose
\begin{eqnarray*} &&\Lambda(\mathbb{R}^{m-1})=\Lambda(\mathbb{R})\otimes...\otimes\Lambda(\mathbb{R})\quad\text{and}\\
&&S=\textstyle\sum_{1\le i\le m-1}\Id\otimes...\otimes\Id\otimes S_i\otimes\Id\otimes...\otimes\Id\quad\text{where}\\
&&S_i=0\text{ on }\Lambda^0(\mathbb{R})\text{ and }S_i=-A_i\text{
on }\Lambda^1(\mathbb{R})\,. \end{eqnarray*}

The supertrace of $\Id$ is zero. Furthermore, the supertrace of
the tensor product is the product of the supertraces. Thus only
$(m-1)!S_1\otimes...\otimes S_{m-1}$ survives in the supertrace of
$S^{m-1}$. Since the supertrace of $S_i$ is $-A_i$, we have that:
\begin{equation}\label{eqn4.f}
\textstyle\textstyle\sum_p(-1)^p\Tr_{\Lambda^p(\mathbb{R}^{m-1})}\{S^{m-1}\}=(m-1)!\mathcal{A}\,.
\end{equation}
Assertion (1) part one now follows from Equations (\ref{eqn4.b}),
(\ref{eqn4.e}), and (\ref{eqn4.f}).

The invariant $a_{m,m,0}^{d+\delta}$ is homogeneous of weight
$m-1$ and is in the kernel of $r$. Thus we can use exactly the
same line of argument to show:
$$a_{m,m,0}^{d+\delta}(0,g)(0)=(m-1)!\mathcal{A}\cc_{m,m,0}^0\,.$$
We use Theorem \ref{thm1.1} to evaluate
$a_{m,m,0}^{d+\delta}(0,g)(0)$ and establish Assertion (1) part
two.

The proof of Assertion (2) is similar. Let $m\ge3$. To simplify
the notation, set $$P_{m+1}(g):=a_{m+1,m,0}^{d+\delta}(0,g),\quad
c_{m+1}:=c_{m+1,m,0}^{3,0},\quad\text{and}\quad\cc_{m+1}:=\cc_{m+1,m,0}^3\,.$$
Let $(u_1,u_2,y_1,...,y_{m-3},r)$ be coordinates on
$\mathbb{R}^m$. Let $f(y)$ satisfy the normalizations of Equation
(\ref{eqn4.a}). We set $M=\{x\in\mathbb{R}^m:r\ge f(y)\}$ and
$$ds^2_M:=du_1^2+e^{-A_0u_1^2r}du_2^2+dy_1^2+...+dy_{m-3}^2+dr^2\,.
$$
Then $R(\cdot)(0)=0$ and the non-vanishing
components of $L$ and $\nabla R$ at the origin are given, up to the usual $\mathbb{Z}_2$ symmetries, by: \begin{eqnarray*} &&L(\partial_i^y,\partial_j^y)(0)=-A_i\delta_{ij},\quad\text{and}\\
&&R(\partial_1^u,\partial_2^u,\partial_2^u,\partial_1^u;\partial_r)
=R(\partial_1^u,\partial_2^u,\partial_2^u,\partial_r;\partial_1^u)=A_0\,.
\end{eqnarray*}
Let $\mathcal{A}=(-1)^{m-3}A_0A_1...A_{m-3}$. We apply Lemma
\ref{lem1.3} to see \begin{equation}\label{eqn4.g}
P_{m+1}(g)(0)=2(m-3)!c_{m+1}\mathcal{A}\,.
\end{equation}
We now let $\KK$ be the ideal generated by the elements
$\{A_0^2,A_1^2,...,A_{m-3}^2\}$. If we set $A_0=0$, then the
manifold is a product of the manifold considered previously with a
flat factor. This shows that $\nabla^kR(0)$, $\nabla^kE(0)$,
$\nabla^k\Omega(0)$ are all divisible by $A_0$ for $k\ge1$ and
vanish if $k=0$.

We consider terms which can give rise to $\mathcal{A}$ after
taking the supertrace. Let $\mathcal{E}$ denote a generic
polynomial in the tangential covariant derivatives of $L$, of $S$,
and of $\chi$ when $A_0$ is set to zero. Since we are not
interested in terms which are divisible by $A_0^2$ and since $A_0$
has weight $3$, we factor out a term which can be linear in $A_0$
to express $P_{m+1}$ symbolically as: \begin{eqnarray*}
P_{m+1}&=&\textstyle\textstyle\sum_p(-1)^p\Tr_{\Lambda^p(M)}\big\{
\textstyle\sum_{k\ge1}\nabla^kR\cdot\mathcal{E}_{m-k-2}^R+\sum_{k\ge1}\nabla^kE\cdot\mathcal{E}_{m-k-2}^E\\
&+&\textstyle\sum_{k\ge1}\nabla^k\Omega\cdot\mathcal{E}_{m-k-2}^\Omega
 +\sum_{k\ge2}\tilde\nabla^kL\cdot\mathcal{E}_{m-k-1}^L\\
&+&\textstyle\sum_{k\ge2}\tilde\nabla^kS\cdot\mathcal{E}_{m-k-1}^S+
 \textstyle\sum_{k\ge3}\tilde\nabla^k\chi\cdot\mathcal{E}_{m-k}^\chi\big\}\,.
\end{eqnarray*}

We set $A_0=0$ in studying the `coefficient' monomials
$\mathcal{E}$. Thus the arguments given above in the proof of
Assertion (1) shows only powers of $S$ are relevant so
\begin{eqnarray}
P_{m+1}&=&\textstyle\textstyle\sum_p(-1)^p\Tr_{\Lambda^p(M)}\big\{\sum_{k\ge1}\nabla^kR\cdot
S^{m-k-2}+\sum_{k\ge1}\nabla^kE\cdot S^{m-k-2}\nonumber\\
&+&\textstyle\sum_{k\ge1}\nabla^k\Omega\cdot S^{m-k-2}
+\sum_{k\ge2}\tilde\nabla^kL\cdot S^{m-k-1}\label{eqn4.h} \\
&+&\textstyle\sum_{k\ge2}\tilde\nabla^kS\cdot S^{m-k-1}
+\textstyle\sum_{k\ge3}\tilde\nabla^k\chi\cdot
S^{m-k}\big\}\,.\nonumber \end{eqnarray}

By Lemma \ref{lem3.7},
\begin{equation}\label{eqn4.i}  \textstyle\textstyle\sum_p(-1)^p\Tr_{\Lambda^p(M)}\{S^k\}=0\quad\text{for}\quad k<m-1\,. \end{equation} Thus the terms in $\nabla^kR$ and $\tilde\nabla^kL$ do not appear in Equation (\ref{eqn4.h}) since, being scalars, they could be moved outside $\Tr$. As $\Omega$ is skew-adjoint and $S$ is self-adjoint, this term does not appear. Terms involving $\tilde\nabla^kS$ must be fully contracted and, modulo lower order terms which can be absorbed at an earlier stage, have the form: $$S_{:a_1a_1a_2a_2...}S^k=\ffrac1{k+1}\{S^{k+1}\}_{:a_1a_1a_2a_2...}+O(A_0^2)\,.$$
Thus by Equation (\ref{eqn4.i}) such terms do not arise in
Equation (\ref{eqn4.h}). A similar argument can be used to
eliminate the terms $\chi_{:a_1a_1a_2a_2...}S^k$ from Equation
(\ref{eqn4.h}).

Extend $S$ to be covariant constant along the geodesic normal rays
from the boundary. This permits us to move covariant derivatives
outside the trace once again. We apply Lemma \ref{lem3.7} to see
$$\textstyle\sum_p(-1)^p\Tr_{\Lambda^p(M)}\{ES^k\}=0\quad\text{for}\quad
k<m-3\,.$$ Thus exactly one covariant derivative of $E$ can appear
and Equation (\ref{eqn4.h}) becomes $$
P_{m+1}(g)(0)=\cc_{m+1}\textstyle\textstyle\sum_p(-1)^p\Tr_{\Lambda^p(\mathbb{R}^m)}\{E_{;m}S^{m-3}\}(0)\,.
$$

If $m=3$, then
$\textstyle\sum_p(-1)^p\Tr_{\Lambda^p(\mathbb{R}^3)}\{E\}=0$. This
implies
$$\textstyle\sum_p(-1)^p\Tr_{\Lambda^p(\mathbb{R}^3)}\{E_{;m}\}=0$$
and hence $c_{m+1}=0$ as desired.

Suppose that $m\ge4$. Since $S$ vanishes on
$\Lambda(\mathbb{R}^{m-1})^\perp$, we have $$
P_{m+1}(g)(0)=\cc_{m+1}\textstyle\textstyle\sum_p(-1)^p\Tr_{\Lambda^p(\mathbb{R}^{m-1})}\{E_{;m}S^{m-3})\}\,
$$
We may decompose $\Lambda(\mathbb{R}^{m-1})=\Lambda(\mathbb{R}^2)\otimes\Lambda(\mathbb{R}^{m-3})$ to express $E_{;m}=\tilde E\otimes\Id$ and $S=\Id\otimes\tilde S$. This then leads to the corresponding decomposition of the supertrace \begin{eqnarray*} &&\textstyle\textstyle\sum_p(-1)^p\Tr_{\Lambda^p\mathbb{R}^{m-1}}\{E_{;m}S^{m-3}\}\\
&=&\textstyle\sum_a(-1)^a\Tr_{\Lambda^a(\mathbb{R}^2)}\{\tilde
E_{;m}\}\cdot\sum_b(-1)^b\Tr_{\Lambda^b(\mathbb{R}^{m-3}}\{\tilde
S^{m-3}\}\,. \end{eqnarray*} The computation performed above shows
that the supertrace of $S^{m-3}$ on $\mathbb{R}^{m-3}$ is
$(-1)^{m-3}(m-3)!A_1....A_{m-3}$. A direct calculation of the
supertrace of $E_{;m}$ on $\mathbb{R}^2$ yields $2A_0$. The final
assertion of Lemma \ref{lem4.2} now follows. \end{proof}

We continue our study by using the various functorial properties
to show:

\begin{lemma}\label{lem4.3}
\begin{enumerate}
\item $\cc_{n,m,k}^i=(4\pi)^{-(m-1)/2}\cc_{n,1,k}^i$.
\item If $n\ge3$, then $\cc_{n,m,1}^0=\frac12\cc_{n,m,0}^0$.
\item If $n\ge5$, then $\cc_{n,m,0}^3=\cc_{n-2,m,1}^0$. \end{enumerate} \end{lemma}

To prove Assertion (1), we use product formulas. Let $M_1=T^{m-1}$
be the torus and let $D_1$ be the scalar Laplacian. Since the
structures are flat,
$$a_{n,m-1}(x_1,D_1)=\left\{\begin{array}{rll}
(4\pi)^{-(m-1)/2}&\text{if}&n=0,\\
0&\text{if}&n>0.\end{array}\right.$$ Let
$(M_2,D_2)=([0,1],-\partial_r^2)$. Let $M=M_1\times M_2$ and
$D=D_1+D_2$. Let $\BB=\nabla_{e_m}+S$ where $S$ is constant and
where $e_m$ is the inward unit normal; $e_m=\partial_r$ when $r=0$
and $e_m=-\partial_ r$ when $r=1$. An analogous argument to that
which was used to establish Lemma \ref{lem2.3} (4) can be used to
establish the following identity from which Assertion (1) follows:
\begin{eqnarray*}
a_{n,m,k}(y,D,\BB)&=&\textstyle\sum_{n_1+n_2=n}a_{n_1,m-1}(x_1,D_1)\cdot
a_{n_2,1,k}(y_2,D_2,\BB)\\
&=&(4\pi)^{-(m-1)/2}a_{n,1,k}(y_2,D_2,\BB).\end{eqnarray*}

In view of Assertion (1), it suffices to take $m=1$ in the proof
of the remaining assertions. We use results from \cite{BG90}. Let
$M:=[0,1]$ and let $D_0:=-\partial_r^2$. We choose $f$ so that $f$
vanishes identically near $r=1$ so only the component $r=0$ where
$\partial_r$ is the inward unit normal is relevant. To prove
Assertion (2), we consider a conformal variation
$D_\varepsilon:=e^{-2\varepsilon f}D_0$. Then: \begin{eqnarray*}
\partial_{\varepsilon}S|_{\varepsilon=0}=-\ffrac12f_{;m}\quad\text{and}\quad
\partial_{\varepsilon}a_n(1,D_\varepsilon)|_{\varepsilon=0}=(1-n)a_n(f,D_0)\,.
\end{eqnarray*}
For $n\ge3$, $f_{;m}S^{n-2}$ arises from no other term. Thus we may show $\cc_{n,1,0}^0=\frac12\cc_{n,1,0}^0$ by computing: \begin{eqnarray*} &&\partial_\varepsilon a_n(1,D_\varepsilon)=\textstyle\partial_\varepsilon\int_{\partial M}\cc_{n,1,0}^0S^{n-1}dy|_{\varepsilon=0}+\ldots\\
&=&-\ffrac12(n-1)\cc_{n,1,0}\textstyle\int_{\partial
M}f_{;m}S^{n-2}dy+\ldots\\
&=&(1-n)a_n(f,D_0)=(1-n)\textstyle\int_{\partial
M}f_{;m}S^{n-2}dy+\ldots\,. \end{eqnarray*}

To prove Assertion (3), we consider a scalar variation
$D_\varrho:=D_0-\varrho f$. We have: $$
\partial_{\varrho}a_n(1,D_\varrho)|_{\varrho=0}=a_{n-2}(f,D_0)\,.
$$
If $n\ge5$, then this is the only way a term involving $f_{;m}S^{n-4}$ can arise. We show $\cc_{n,1,0}^3=\cc_{n-2,1,1}^0$ by computing: \begin{eqnarray*} &&\partial_{\varrho}a_n(1,D_\varrho)|_{\varrho=0}=\textstyle\partial_\varepsilon\int_{\partial M}\cc_{n,1,0}^3E_{;m}S^{n-4}dy|_{\varrho=0}+\ldots\\
&=&a_{n-2}(f,D_0)=\textstyle\int_{\partial
M}\cc_{n-2,1,1}^0f_{;m}S^{n-4}dy+\ldots\,.\qquad\qedbox
\end{eqnarray*}

\medbreak\noindent{\bf Remark:} Lemma \ref{lem4.3} (2) fails if
$n=2$ and Lemma \ref{lem4.3} (3) fails if $n=4$ as there are
interior terms which also contribute to the variational formulae.

\section{Proof of Theorem \ref{thm1.2}}\label{Sect5} We
use Lemmas \ref{lem4.1}, \ref{lem4.2}, and \ref{lem4.3} to determine the constants of Lemma \ref{lem1.3}: \begin{eqnarray*} c_{m+1,m}&=&\ffrac1{\sqrt\pi8^{\bar m}\pi^{\bar m}\bar m!}\quad\text{for}\quad m=2\bar m+1,\\ c_{m+1,m,1}^k&=&\ffrac1{\pi^k8^kk!}c_{m-2k+1,m-2k,1}^0=\ffrac1{\pi^k8^kk!}\cc_{m-2k+1,m-2k,1}^0\\
 &=&\ffrac1{2\pi^k8^kk!}\cc_{m-2k+1,m-2k,0}^0=\ffrac{2\sqrt\pi}{2\pi^k8^kk!}\cc_{m-2k+1,m-2k+1,0}^0\\
 &=&\ffrac{\sqrt\pi}{8^k\pi^kk!\operatorname{vol}(S^{m-2k})(m-2k)!},\\
c_{m+1,m,0}^{1,k}&=&\ffrac1{\pi^k8^kk!}c_{m+1-2k,m-2k,0}^{1,0}
 =\ffrac1{\sqrt\pi\pi^k8^kk!\operatorname{vol}(S^{m-2k-2})(m-2k-2)!},\\
c_{m+1,m,0}^{2,k}&=&\ffrac1{\pi^k8^kk!}c_{m+1-2k,m-2k,0}^{2,0}=0,\\
c_{m+1,m,0}^{3,k}&=&\ffrac1{\pi^k8^kk!}c_{4,3,0}^{3,0}=0\quad\text{for}\quad 2k=m-3,\\ c_{m+1,m,0}^{3,k}&=&\ffrac1{\pi^k8^kk!}c_{m-2k+1,m-2k,0}^{3,0}=\ffrac1{\pi^k8^kk!}\cc_{m-2k+1,m-2k,0}^3\\
&=&\ffrac1{\pi^k8^kk!}\cc_{m-2k-1,m-2k,0}^0=\ffrac1{2\sqrt\pi\pi^k8^kk!}\cc_{m-2k-1,m-2k-1,0}^0\\
&=&\ffrac1{2\sqrt\pi\pi^k8^kk!}c_{m-2k-1,m-2k-1,0}^0\\
&=&\ffrac1{2\sqrt\pi\pi^k8^kk!\operatorname{vol}(S^{m-2k-2})(m-2k-2)!}\quad\text{for}\quad
2k<m-3 \,.\quad \qedbox \end{eqnarray*} 
\end{document}